# Nonlinear friction in quantum mechanics

Roumen Tsekov
Department of Physical Chemistry, University of Sofia, 1164 Sofia, Bulgaria

The effect of nonlinear friction forces in quantum mechanics is studied via dissipative Madelung quantum hydrodynamics. A new thermo-quantum diffusion equation is derived, which is solved for the particular case of quantum Brownian motion with cubic friction. It is extended by a chemical reaction term to describe quantum reaction-diffusion systems with nonlinear friction as well.

Nonlinear friction forces are a problem in classical Brownian motion for a long time [1]. They are described either by Langevin or by Fokker-Planck equations [2-4]. However, the rigorous generalized Langevin equation is linear, which points out that nonlinear friction forces possess a macroscopic hydrodynamic origin [1]. For this reason, they are not present in the modern quantum theory of open systems [5]. The scope of the present paper is to investigate the nonlinear friction effect on quantum mechanics. The analysis is based on a dissipative Madelung quantum hydrodynamics. Thus, the nonlinear friction is a result of nonlinear non-equilibrium thermodynamics in contrast to the traditional mechanical nonlinear friction forces.

A quantum particle in vacuum is described by the Schrödinger equation

$$i\hbar \partial_t \psi = -\hbar^2 \partial_x^2 \psi / 2m + U\psi \qquad (1)$$

where $m$ is the particle mass and $U$ is an external potential. The complex wave function can be generally presented in its polar form $\psi = \sqrt{\rho} \exp(iS/\hbar)$, where $\rho$ is the probability density and $S/\hbar$ is the wave function phase. Introducing this Madelung presentation in Eq. (1) results in the following two equations [6]

$$\partial_t \rho = -\partial_x (\rho V) \qquad m\partial_t V + mV\partial_x V = -\partial_x p_Q / \rho - \partial_x U \qquad (2)$$

corresponding to the imaginary and real parts, respectively. The first equation is a continuity one with $V \equiv \partial_x S / m$ being the hydrodynamic-like velocity in the probability space. The second equation is a macroscopic force balance, where the quantum effects are solely included in the quantum pressure $p_Q \equiv -(\hbar^2/4m)\rho \partial_x^2 \ln \rho$. Note that the latter depends both on the local density and its spatial derivatives and, hence, the Madelung hydrodynamics is a non-local theory.

The Madelung presentation of the Schrödinger equation opens a door for introduction of dissipative forces in quantum mechanics. The friction force of a particle in a classical environment depends naturally on the particle velocity. Hence, one can add a macroscopic friction force $f(V)$ in the force balance (2) to obtain

$$m\partial_t V + mV\partial_x V = -\partial_x(p_Q + k_B T\rho)/\rho - \partial_x U + f(V) \tag{3}$$

Here the new pressure term accounts for the osmotic thermal pressure due to the environment temperature $T$. Thus one arrives to a dissipative Madelung hydrodynamics. At strong friction the inertial terms on the left-hand-site of Eq. (3) can be neglected as compared to the friction force and the hydrodynamic-like velocity can be expressed in the form $V = f^{-1}(\partial_x \mu)$, where $f^{-1}$ is the inverse function of $f$ and $\mu \equiv Q + k_B T \ln \rho + U$ is the local chemical potential. The chemical potential term $Q \equiv -\hbar^2 \partial_x^2 \sqrt{\rho}/2m\sqrt{\rho}$, corresponding to the quantum pressure via the Gibbs-Duhem relation $dp_Q = \rho dQ$, is in fact the Bohm quantum potential. While the latter is an icon in the de Broglie-Bohm theory, the symbol of the Madelung hydrodynamics is $p_Q$. Introducing now this expression for $V$ into the continuity equation (2) results in a generalized nonlinear diffusion equation

$$\partial_t \rho = -\partial_x[\rho f^{-1}(\partial_x \mu)] \tag{4}$$

The equilibrium solution of Eq. (4) corresponds to $V = 0$ or a constant chemical potential, which is in accordance to the rules of thermodynamics. A typical model for $f^{-1}$ in activated diffusive processes is the hyperbolic sine function following from the Arrhenius law.

Equation (4) is valid for arbitrary friction forces. Usually the friction force is well approximated by the expression $f(V) = -b_1 V - b_3 V^3$ with two friction coefficients, a linear one $b_1$ and a cubic one $b_3$ [7]. At low hydrodynamic velocity not very far from the equilibrium the cubic term becomes negligible and, hence, $f^{-1}(\partial_x \mu) = -\partial_x \mu / b_1$. Thus Eq. (4) reduces to a quantum Smoluchowski equation [8]

$$\partial_t \rho = \partial_x[\rho \partial_x(Q + U)/b_1 + D\partial_x \rho] \tag{5}$$

where $D = k_B T/b_1$ is the classical Einstein diffusion constant. The solution of Eq. (5) for a free particle at zero temperature is a Gaussian distribution density with dispersion obeying the sub-diffusive quantum law $\sigma^2 = \hbar\sqrt{t/mb_1}$ [8]. In the opposite case far from equilibrium the cubic term dominates the friction force and $f^{-1}(\partial_x \mu) = -\sqrt[3]{\partial_x \mu / b_3}$. Thus Eq. (4) acquires the following strongly nonlinear form

$$\partial_t \rho = \partial_x [\rho \sqrt[3]{\partial_x (Q + k_B T \ln \rho + U)/b_3}] \tag{6}$$

For a classical particle moving in a biquadratic external potential $U = Kx^4/4$ the solution of Eq. (6) reads $\rho = \Gamma(3/4)\exp(-x^4/4\sigma^4)/\pi\sigma$, where the average displacement evolves in time according to the equation

$$F(1/3,1/3;4/3;K\sigma^4/k_B T)\sqrt[3]{K\sigma^4/k_B T} = (4/3)\sqrt[3]{K/b_3}t \tag{7}$$

Here $\Gamma$ and $F$ are the gamma and hypergeometric functions, respectively. The plot of Eq. (7) is shown in Fig. 1. As is seen, initially the evolution is super-diffusive, than passes through a normal diffusive regime and ends with a sub-diffusive part. At infinite time $\sigma_\infty^4 = k_B T/K$ and the probability density reduces to the equilibrium Boltzmann distribution. In the case of a free classical particle with a cubic friction force Eq. (7) provides a super-diffusive classical law $\sigma^2 = \sqrt{64 k_B T t^3 / 27 b_3}$. Hence, the nonlinear friction accelerates the particle diffusion, which is, however, non-Gaussian.

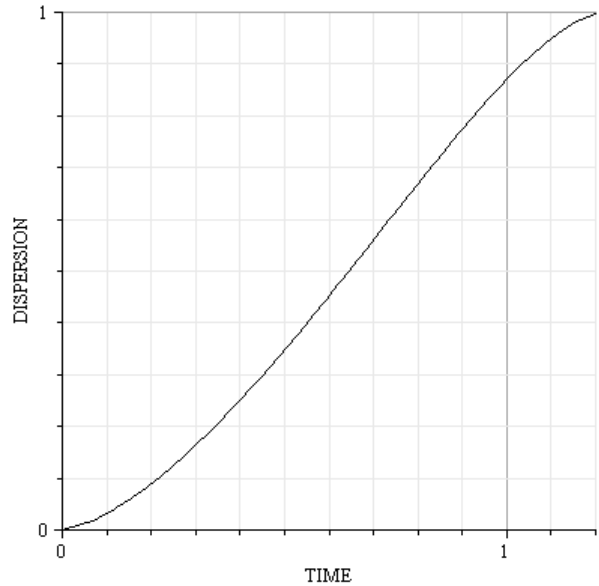

**Fig. 1** Dimensionless dispersion $\sqrt{K/k_B T}\sigma^2$ vs. dimensionless time $(4/3)\sqrt[3]{K/b_3}t$.

In the case of free quantum diffusion at zero temperature Eq. (6) reduces to

$$\partial_t \rho = \partial_x(\rho \sqrt[3]{\partial_x Q / b_3}) = -\partial_x[\rho \sqrt[3]{\hbar^2(\partial_x \ln \rho \partial_x^2 \ln \rho + \partial_x^3 \ln \rho)/4mb_3}] \tag{8}$$

At large $x$ one can neglect the third-derivative term in the brackets of Eq. (8) and the solution of the remaining equation is $\rho = 3\sqrt[6]{3}\Gamma(2/3)\exp(-|x|^3/3\sigma^3)/4\pi\sigma$. Surprisingly, the corresponding displacement obeys a normal diffusive law $\sigma^2 = 2\sqrt[3]{\hbar^2/2mb_3}\,t$. This unexpected result shows that the quantum sub-diffusivity compensate the super-diffusivity originating from the cubic friction in such a way that the final result corresponds formally to the classical Einstein law with a novel quantum diffusion constant $\sqrt[3]{\hbar^2/2mb_3}$. The distribution density above is, however, non-Gaussian again.

Generally, it is possible to find the inverse function of the complete nonlinear friction force $f(V) = -b_1 V - b_3 V^3$ and to perform the corresponding analysis of Eq. (4). The physical transparency will suffer, however, due to mathematical complications. We expect the appearance of many sub- and super-diffusive regimes, which alternatively can be formally described via fractional diffusion equations [9]. In the case of diffusion in a structured environment it is expected that the friction coefficients $b_1$ and $b_3$ will depend on the local particle position $x$ [10]. This will not change, however, the validity of the general diffusion equation (4). Moreover, any more advanced model for the local chemical potential $\mu$ could be directly employed in Eq. (4). In this way, for instance, a relativistic correction on the nonlinear quantum Brownian motion can be obtained [11].

Thermodynamically, Eq. (4) can be also accomplished by a chemical reaction rate $R$ to describe quantum reaction-diffusion systems as well [12]

$$\partial_t \rho = -\partial_x[\rho f^{-1}(\partial_x \mu)] + R(\rho) \tag{9}$$

This nonlinear equation accounts for the nonlinear friction and quantum effects on the typically nonlinear temporal and spatial reactive self-organizations. In the case of a linear friction model without an external potential Eq. (9) reduces to

$$\partial_t \rho = \partial_x(\rho \partial_x Q / b_1) + D\partial_x^2 \rho + R \approx -\hbar^2 \partial_x^4 \rho / 4mb_1 + D\partial_x^2 \rho - k(\rho - \rho_{eq}) \tag{10}$$

where the last approximate expression represents a linearization around the constant equilibrium density $\rho_{eq}$, defined by $R(\rho_{eq}) = 0$, and $k \equiv -(\partial_\rho R)_{eq}$ is a first-order reaction rate constant. The Fourier image of the solution of Eq. (10) reads

$$\rho_q = \rho_{eq}[1-\exp(-kt)]\delta(q) + \exp[-(k+Dq^2+\hbar^2 q^4/4mb_1)t] \tag{11}$$

where $q$ is the wave vector. As is seen, the quantum term describes a biharmonic diffusive process. Thus, quantum effects result in a faster spreading of the initially localized wave packet. It dominates at low temperature, where the classical diffusion constant $D$ becomes negligible.

In the case of lack of friction one can add the reaction rate directly into Eq. (2) to obtain

$$\partial_t \rho = -\partial_x(\rho V) + R(\rho) \qquad \partial_t V + V\partial_x V = -\partial_x(U+Q)/m \tag{12}$$

If the velocity is nearly quasi-static $V\partial_x V \gg \partial_t V$ an integration of the second dynamic equation yields the hydrodynamic-like velocity $V^2 = V_0^2 - 2(U+Q)/m$, where $V_0$ is an integration constant. Introducing this expression in the first equation leads to

$$\partial_t \rho = -\partial_x[\rho\sqrt{V_0^2 - 2(U+Q)/m}] + R(\rho) \tag{13}$$

As is seen, this is a strongly nonlinear equation even if the friction is missing. In the case of a free particle ($U=0$) after linearization around the equilibrium density Eq. (13) reduces to

$$\partial_t \rho = -V_0 \partial_x \rho - \hbar^2 \partial_x^3 \rho / 4m^2 V_0 - k(\rho - \rho_{eq}) \tag{14}$$

This equation describes a quantum diffusion of the convective flow. The Fourier image of the solution of Eq. (14) reads

$$\rho_q = \rho_{eq}[1-\exp(-kt)]\delta(q) + \exp[-(k+iqV_0 - iq^3\hbar^2/4m^2 V_0)t] \tag{15}$$

Thus, the effective convection rate $V_0[1-(\hbar q/2mV_0)^2]$ possesses a quantum retardation effect, which scales with the convective de Broglie wave length $\hbar/2mV_0$.

The equations above describe chemical binding of an electron by the environment. In some cases electrons interact effectively each other; an example is a Cooper pair. To describe the dissipation in a gas of electrons one could employ the concept of viscous friction among them. Hence, in a Stokes flow at zero temperature of an electron gas with kinematic viscosity $\nu$ the pressure force balance (3) reads $\partial_x p_Q = f(V)\rho = \partial_x(m\rho\nu\partial_x V)$. The integration of this equation is straightforward and yields an expression for the hydrodynamic-like velocity

$$V = -(\hbar^2/4m^2\nu)\partial_x \ln\rho \tag{16}$$

Introducing now Eq. (16) in the reactive continuity equation the latter changes to

$$\partial_t\rho = -\partial_x(\rho V) + R(\rho) = (\hbar^2/4m^2\nu)\partial_x^2\rho + R(\rho) \tag{17}$$

As is seen, Eq. (17) formally coincides with the classical reaction-diffusion equation and the effective diffusion constant reads $\hbar^2/4m^2\nu$ [13]. Since in dilute gasses the self-diffusion constant equals to the kinematic viscosity, it follows from $\hbar^2/4m^2\nu = \nu$ that the kinematic viscosity of an electron gas at zero temperature is simply given by $\nu = \hbar/2m$. The corresponding dynamic viscosity $m\rho\nu = \rho\hbar/2$ correlates well to a recent report that at low temperature the dynamic viscosity of a unitary Fermi gas scales universally with the density [14]. This is not surprising, however, since the collisions between electrons are subject of the Heisenberg uncertainty principle. Hence, the product of the mean free path and root mean square momentum of a particle is limited from below by the half of the Planck constant.

Finally, the measurement problem in quantum mechanics can be described also as an irreversible nonlinear process. Obviously, the observer receives information about the quantum system via the measurement, being equivalent to exchange of informational entropy. Usually in quantum mechanics the measurement details are not specified but it is clear that this process disturbs the initially isolated quantum system via additional terms in the system Hamiltonian. A plausible assumption is that the entropy contributes additively to the system Hamiltonian via the information density $I$ [15]. In this case, the Schrödinger equation (1) acquires during the measurement process the following form

$$i\hbar\partial_t\psi = -\hbar^2\partial_x^2\psi/2m + U\psi - I\psi \tag{18}$$

It is important to note here that the information density $I$ depends on the wave function via the probability density. For instance, $I$ is proportional to $-\ln\rho$ or $1-\rho$ for the Shannon or linear entropy, respectively. Thus, Eq. (18) is a nonlinear Schrödinger equation and, hence, linear combination solutions and the quantum superposition principle are no more valid. Thus, the initial wave function superposition collapses in the particular state corresponding to the measured value [16]. Since the information density $I$ is a projective function of $\psi$ any further measurement of the same quantity will deliver the same result, which corresponds exactly to the von Neumann measurements.

How it is well known, the kinetic energy term in the Schrödinger equation represents, in fact, the Fisher information density [17-20]. Therefore, Eq. (18) describes an information exchange between the internal Fisher and external observer entropies during the measurement process, adjusted correspondingly by the interaction potential $U$. It is known [20] that the pro-duction of Boltzmann entropy via diffusive processes equals to the Fisher information. The hydrodynamic form of Eq. (18) reads

$$\partial_t \rho = -\partial_x(\rho V) \qquad m\partial_t V + mV\partial_x V = -\partial_x p_Q/\rho + \partial_x I - \partial_x U \qquad (19)$$

Comparing how this result with Eq. (3) shows that the thermal osmotic pressure $k_B T \rho$ is equivalent to the Shannon entropy density multiplied by temperature, $I = -k_B T \ln \rho$. Hence, the effect of a thermal environment consists in continuous measurements as well. If the communication between the system and environment is imperfect, a dissipative force $f(V)$ should also appear in Eq. (19) to describe disinformation by the environment due to the Second Law.